\documentstyle[prl,aps]{revtex} 
\begin{document}

\draft
\def\x{{\bf x}}
\def\y{{\bf y}}
\def\R{{\bf R}}

\font\large=cmbx10 at 12 pt  
\newcount\equationno      \equationno=0
\newtoks\chapterno \xdef\chapterno{}
\def\eqn{\eqno\eqname}
\def\eqname#1{\global \advance \equationno by 1 \relax
\xdef#1{{\noexpand{\rm}(\chapterno\number\equationno)}}#1}

\def\la{\mathrel{\mathchoice {\vcenter{\offinterlineskip\halign{\hfil
$\displaystyle##$\hfil\cr<\cr\sim\cr}}}
{\vcenter{\offinterlineskip\halign{\hfil$\textstyle##$\hfil\cr<\cr\sim\cr}}}
{\vcenter{\offinterlineskip\halign{\hfil$\scriptstyle##$\hfil\cr<\cr\sim\cr}}}
{\vcenter{\offinterlineskip\halign{\hfil$\scriptscriptstyle##$\hfil\cr<\cr\sim\cr}}}}}

\def\ga{\mathrel{\mathchoice {\vcenter{\offinterlineskip\halign{\hfil
$\displaystyle##$\hfil\cr>\cr\sim\cr}}}
{\vcenter{\offinterlineskip\halign{\hfil$\textstyle##$\hfil\cr>\cr\sim\cr}}}
{\vcenter{\offinterlineskip\halign{\hfil$\scriptstyle##$\hfil\cr>\cr\sim\cr}}}
{\vcenter{\offinterlineskip\halign{\hfil$\scriptscriptstyle##$\hfil\cr>\cr\sim\cr}}}}}
\reversemarginpar


\title{ Quantum structure of spacetime and blackhole entropy}
\author{T. Padmanabhan\thanks{paddy@iucaa.ernet.in}}
\address{IUCAA, Post Bag 4, Ganeshkhind, Pune 411 007, India.\\
IUCAA preprint 1/98 - January, 1998}
\maketitle

\begin{abstract}
The gap between a microscopic  theory for quantum spacetime and 
the semiclassical physics of blackholes  is bridged by treating 
the blackhole spacetimes as highly excited states of a class of 
 {\it nonlocal} field theories. 
All the blackhole thermodynamics is shown to arise from 
asymptotic form of the dispersion relation 
satisfied by the elementary excitations 
of these field theories. These models involve, quite generically,
 fields which are: (i) smeared over
 regions of the order of Planck length and (ii) possess correlation
functions which have universal short distance behaviour.  
\end{abstract}

\vskip 0.5 in
\noindent
Simple thought experiments  suggest that
there exists an operational limitation in measuring lengths (and times)
smaller than the Planck length $L_P\equiv (G\hbar/c^3)^{1/2}$ (see
for example, ref. [1]). The correct theory of quantum gravity should 
incorporate this limitation in a natural manner just as the correct
quantum mechanical theory (based on non commuting operators) incorporates
the uncertainty principle. String theories  as well as the spin network 
formalism based on Ashtekar variables do seem to have ingredients 
to describe the quantum microstructure of spacetime in such a manner
that $L_P$ arises as a natural lower bound to lengthscales [2,3]. In these
approaches, the spacetime continuum arises as a nonperturbative quantum condensate
of the basic variables. The description of spacetime geometry as a solution 
to Einstien equation is analogous to the description of, say, a gaseous system
by macroscopic parameters through an equation of state. Given the 
microscopic theory, there should exist certain well defined procedure for
obtaining the continuum description.

Unfortunately, we do not yet have a clearly spelt out quantum 
theory of gravity. It is therefore important to ask whether one could 
obtain some general features of such a theory based on our knowledge 
of the macroscopic description. It is obvious that
such an ``inverse" process --- which is similar to an attempt
in understanding the quantum nature of radiation field, given the macroscopic
description of the blackbody radiation ---  will not possess a unique solution; however, some key features, which must be respected by any underlying quantum description of 
spacetime, should arise from such an analysis. 

Since quantum structure of spacetime is likely to reveal itself only at 
energies close to Planck energy, $E_P = L_P^{-1}$, its effects are most likely to influence
description of virtual processes taking place at Planck energies and above.
In general, such virtual processes cannot be described without 
knowing the details of the theory. One fortunate exception occurs in classical
spacetimes containing compact infinte redshift surfaces. I  give 
detailed arguments elsewhere [4] to suggest that blackhole evaporation can be thought of as the deexcitation of degrees of freedom located within a 
few Planck lengths from the event horizon. The existence of infinite 
redshift allows virtual processes at super Planckian energies to 
manifest as phenomena at sub Planckian energies which are describable in 
terms of semiclassical continuum physics. 
It follows that the existence of blackhole evaporation will put
certain constraints on the correct description of quantum microstructure though
it is still likely to leave fair amount of liberty as regards the details [5].

 I  show below that it is possible to construct toy effective 
field theories which correctly describe the semi classical thermodynamics
of blackholes. This analysis  helps to delineate those features
of the effective field theories which are essential to reproduce the 
correct behaviour and reveals the tremendous amount of freedom which still
exists as regards the microscopic description.

Let the quantum microstructure of spacetime be described by certain 
degrees of freedom $q_A$ and energy levels $\epsilon_j$. Classical,
asymptotically flat, spacetimes with mass-energy $M\gg E_P$ will be
made of large number of degrees of freedom combining together in
a coherent manner. Normally, these elementary degrees of freedom
will remain in their ground state. An exception occurs in the
description of classical spacetimes possessing a compact, infinite redshift
surface. In such a case, highly excited states of the basic
degrees of freedom can be populated, at least around the event horizon, 
due to the presence of virtual excitations of arbitrarily high energies.
(The role of infinite redshift surfaces, which distinguishes
a star of mass $M$ from a blackhole of mass $M$, will be elaborated in ref. 4; the
details are irrelevant for this paper).
A blackhole 
spacetime of mass $M$ will correspond to a {\it highly excited} quantum state
of $q_A$'s such that the mean energy of the state $\bar E = M$ with
$M \gg E_P$. 

For a description of such a state, with the least amount of 
additional assumptions, I will rely on pure combinotrics used in 
conventional statistical mechanics. In such a description the mean 
energy can be written in the form
\begin{equation}
\bar E(\beta) = - {\partial\over \partial \beta} \ln Z(\beta); \quad Z(\beta) = \sum_j g_j e^{-\beta \epsilon_j} \cong \int_{\cal C} \rho(z) e^{-\beta z} dz
\label{bhone}\end{equation}
where $Z(\beta)$ is the partition function and $\rho(z)$ is the density 
of states of the system. The second equality for $Z(\beta)$ assumes
the validity of continuum description but keeps the contour of integration ${\cal C}$
in complex plane unspecified; this is convenient since I have to deal with 
density of states which are unbounded along the real axis. Given
the form of $\bar E(\beta)$ one can determine $Z(\beta)$ and, 
by inverting the relation between $Z$ and $\rho$, obtain $\rho(z)$.
If the quantum state describes a semiclassical blackhole, then we must
have $M = \bar E(\beta) = (E_P^2\beta/8\pi)$; this gives $Z(\beta) = \exp(-\beta^2 E_P^2/16 \pi)$. Choosing the contour of integration ${\cal C}$ along the 
imaginary axis, we get $\rho(iy/E_P) = \exp[-4\pi (y/E_P)^2]$. This 
corresponds to the density of states for real energies given by
$\rho(\bar E) = \exp[4\pi (\bar E/E_P)^2]$ for $\bar E \gg E_P$. The corresponding 
entropy is $S(\bar E) \equiv \ln \rho(\bar E) = 4\pi (\bar E/E_P)^2 = {\cal A}/4L_P^2$
where ${\cal A}$ is the area of the event horizon. 
Given a system with this form of density
of states and a mechanism to populate
the excited states as required in the statistical description, one
can reproduce the standard thermodynamics of blackholes.
Hence any description of quantum microstructure
which, in the continuum limit, leads to a density of states of the 
form $\rho(E) \approx \exp[4\pi (\bar E/E_P)^2]$ will correctly reproduce the 
results of blackhole thermodynamics (Populating the excited states is
possible in blackhole spacetimes due to the existence of the infinite
redshift surface.) In fact, this density of state
can only be trusted for $\bar E \gg E_P$. In general, the density of 
states can have a form 
\begin{equation}
\rho(\bar E) \approx \exp[4\pi (\bar E/E_P)^2+{\cal O}(\ln(\bar E/E_P) \cdots]\label{eqn:qrhoneeded}
\end{equation}
 where the leading log corrections are unimportant
for $\bar E\gg E_P$. 

I will now show how several toy field theoretical 
models can be constructed, which have such a density of state. While this 
shows that blackhole entropy, by itself,  cannot provide more information 
regarding the quantum microstructure, it will reveal one essential feature:
{\it All these effective field theory models are nonlocal and involve smearing of the 
fields over a distance of the order of Planck length.} This appears to be 
the key ingredient which is needed to produce the correct continuum limit.

Let us assume that the transition to continuum limit can be described
in terms of certain fields  $\phi(t, {\bf x})$  which are  to be constructed
in some suitable manner from the fundamental variables $q_A$. 
(This process is analogous to obtaining the fluid description of 
a gaseous system made of discrete molecules; in addition, even $t$ and ${\bf x}$ have
to arise  in terms of the microscopic variables in some, as yet unknown, 
fashion.) I take the Lagrangian describing the effective field 
$\phi(t,{\bf x})$ to be 

\begin{equation} L = {1\over 2} \int d^D{\bf x} \, \dot \phi^2 - {1\over 2} \int d^D{\bf x} d^D{\bf y}\, \phi({\bf x}) F({\bf x} - {\bf y}) \phi({\bf y}) = \int {d^D{\bf k}\over (2\pi)^D} {1\over2} \left[ |\dot Q_{\bf k}|^2 - \omega_{\bf k}^2 |Q_{\bf k}|^2\right]\label{eqn:qlag}
\end{equation}
The Lagrangian is non local in the space coordinates ${\bf x}$ which is taken to be $D-$dimensional; the corresponding Fourier space coordinates are labelled
by ${\bf k}$. The quadratic non locality allows us to describe the system
in terms of free harmonic oscillators with a dispersion relation $\omega({\bf k})$ related to $F({\bf r}) $ by 

\begin{equation} 
\omega^2({\bf k}) = \int d^D{\bf r} \, F({\bf r}) e^{i{\bf k}\cdot{\bf r}}
\end{equation}
The energy levels of the system are built out of elementary excitations
with energy $\hbar \omega({\bf k})$ and the partition function 
is given by
\begin{equation}
 Z(\beta) \cong \int {d^D{\bf k}\over (2\pi)^D} \exp[-\beta\omega({\bf k})] = \int dE\, \rho(E) e^{-\beta E}
\end{equation}
where $\rho(E)$ is the Jacobian $\rho(E) = |d^D{\bf k}/dE|$. It is 
straightforward to see that if the dispersion relation $\omega({\bf k})$ has
the asymptotic form 

\begin{equation}
\omega^2({\bf k}) \to {E_P^2 D\over 8\pi} \ln k^2 \qquad ({\rm for}\ k^2 \gg E_P^2)\label{eqn:qasydis}
\end{equation}
then the density of states has the asymptotic form as given in  (\ref{eqn:qrhoneeded}).
Thus a wide class of toy effective field theories of very simple 
nature can reproduce the blackhole thermodynamics. The only nontrivial 
ingredient, compared to conventional field theories, is the non locality in
the ${\bf x}-$space. 

I will now illustrate the above phenomena using the simplest possible choice,
corresponding to $D=1$ and a dispersion relation 
\begin{equation}
\omega^2(k) = {E_P^2\over 8\pi} \ln \left( 1 + {8\pi k^2\over E_P^2}\right) \label{eqn:qdispch}
\end{equation}
The form of $\omega(k)$ for $ k \ll E_P$ is somewhat arbitrary. I 
have chosen it so as to reduce the model to a local, massless, free field theory (with $\omega^2 =  k^2$) in this limit. 
Since quantum gravitational effects should be negligible for $E_P \to \infty$, 
this appears reasonable. 
(In addition to simplicity,
this form can be motivated by other considerations [4].) The density of 
states corresponding to this dispersion relation is given by 

\begin{equation}
 \rho(\bar E) \cong  \exp\left[ 4\pi {\bar E^2\over E^2_P} + {\cal O} \left( \ln {\bar E\over E_P} \right) \right] \equiv \exp S(\bar E) 
\end{equation}
The corresponding blackhole temperature is 

\begin{equation}
 T(\bar E) = \left( {\partial S\over \partial \bar E}\right)^{-1} = {E_P^2 \over 8\pi \bar E} \left[ 1 + {\cal O} \left({ E_P^2\over \bar E^2}\right) \right] \cong {E_P^2\over 8\pi M} 
\end{equation}
for $\bar E = M \gg E_P$. The function $F( r)$ corresponding to the $\omega^2( k)$ in equation (\ref{eqn:qdispch})  is 

\begin{equation}
F(x) = {E_P^2\over 8\pi}\int^\infty_{-\infty} {dk\over 2\pi} e^{-ikx} \ln \left( 1 + {8\pi k^2\over E_P^2}\right) = -{E_P^3\over 8\pi} \left( {L_P\over |x|}\right) \exp \left( - {|x|\over \sqrt{8\pi} L_P}\right)
\end{equation}
for finite, nonzero, $x$. [The logarithmic singularity in the $ k$ integration can 
be handled by standard regularisation techniques; e.g by using the
integral representation of the ln function.] When $L_P \to 0$, the function $F(x)$ is
proportional to the second derivative of Dirac delta function as can be 
seen from the fact that, as $L_P \to 0, \omega^2(k) \to k^2$. In this 
limit, we recover the standard local field theory. 

The functional form of $F(x)$ clearly illustrates the smearing of the 
fields over a region with correlation length $\sqrt{8\pi}\, L_P$. It
can be shown, with more tedious algebra, that this is a generic feature
of the dispersion relation (\ref{eqn:qdispch}) in any dimension $D$. The 
asymptotic structure in (\ref{eqn:qasydis}) governs the short distance behaviour of
$F({\bf x})$ and in $D-$dimension $F({\bf x}) \propto (|{\bf x}|/L_P)^{-D}$ as $|{\bf x}|\to 0$.
In fact, this is the only feature  of $F({\bf x})$ which is needed 
to reproduce the  density of states leading to the correct theory 
of blackhole thermodynamics. 

The physical content of the above analysis can be viewed as follows.
I start with certain loosely defined dynamical variables $q_A$ describing
the quantum microstructure of the spacetime. The dynamical theory
describing $q_A$'s must lead, in suitable limit, to a continuum
spacetime with quantum states having mean energies much larger
than $E_P$. Among them are the classical spacetimes
with compact, infinite redshift surfaces like that of a blackhole forming out of a 
stellar collapse. I describe these blackhole spacetimes
in terms of an intermediate effective field theory in $(D+1)$ dimension. The existence of an infinite redshift surface 
allows the elementary excitations of this field with arbitrarily
high energies to occur in such spacetimes. (In the absence of infinite redshift surface,
one cannot populate high energy states of the toy field so as to
obtain a thermodynamic description). 
Such a theory is nonlocal in space and is based on smearing of fields
over a correlation length of the order of $L_P$. 

In fact, any field theory described by the Lagrangian of the form in (\ref{eqn:qlag})
can be expressed in terms of a free field $\psi$ such that 
$\phi $ is obtained by smearing $\psi$ using a window function $W$; in 
Fourier space, $\psi_{\bf k} = \phi_{\bf k} W_{\bf k}$ with
$|W_{\bf k}|^2 = k^2 F_{\bf k}^{-1}$.
The short distance 
behaviour of such a correlation function is universal and is of the 
form $(|{\bf x}|/L_P)^{-D}$ in $D-$dimension. The blackhole spacetimes are 
interpreted as highly excited states of such a toy field theory which 
itself is built out of more fundamental, and as yet unknown,
variables $q_j$. The semiclassical, thermodynamic behaviour of blackholes can come from such an effective field theory.  

The above analysis 
shows that there is nothing mysterious in completely different
microscopic models (like those based on strings or Ashtekar variables)
leading to similar results regarding blackhole entropy [6]. Any theory 
which has the correct density of states can do this; in fact, the 
models I have described are only a very specific subset of several such
 field theories which can be constructed. The situation is reminiscent
of one's attempt to understand the quantum nature of light
from blackbody radiation. The spectral form of blackbody radiation 
can be derived from the assumption that $E=\hbar \omega$ and is 
quite independent of the details of quantum dynamics of the electromagnetic 
field. Similarly, the blackhole thermodynamics can be explained if 
one treats spacetimes with event horizons as highly excited states of a 
nonlocal field theory whose elementary excitations obey a dispersion 
relation with the asymptotic form given by (\ref{eqn:qasydis}). Within the spirit of 
the current analysis, one need not even identify the $(D+1)$ dimensional
space
as a superset of conventional spacetime. The ${\bf x}$ and $t$ could
represent variables in some abstract space and the spacetime structure
could emerge in a more complicated manner in terms of the fields themselves.
Because of this reason, I have not bothered  about the Lorentz
invariance or other internal symmetries of the toy model. 

While it may not be possible to obtain a unique quantum description of 
spacetime from our knowledge of semiclassical blackhole physics, it 
does give three clear pointers. First is the indirect, but essential, role played by the infinite
redshift surface: It is the existence of such a surface which distinguishes
the {\it star} of mass $M$ from a {\it blackhole} of mass $M$. A stellar spactime will
not be able to populate high energy states of the toy field as required
by the statistical description; in a blackhole spacetime, virtual modes of
arbitarily high energies near the event horizon will allow this to occur. 
(It may be possible to model such a process by studying the interaction of
this toy field with a more conventional field near the event horizon.)
Second one is the asymptotic form of the 
dispersion relation for the elementary excitations which leads to the correct density of states. The third
is the fact that such a dispersion relation almost invariably
leads to smearing of local fields over regions of the order of Planck
length. 

I thank Apoorva Patel and Sayan Kar for several stimulating discussions.

\end{document}